\documentclass[twocolumn,preprintnumbers,amsmath,aps]{revtex4}
\usepackage{graphicx}
\usepackage{dcolumn}
\usepackage{bm}
\usepackage{color}
\usepackage{multirow}
\begin{document}
\def\eq#1{\hspace{1mm}(\ref{#1})}
\def\fig#1{\hspace{1mm}\ref{#1}}
\def\tab#1{\hspace{1mm}\ref{#1}}
\title{Chaotic evolution of the energy of the electron orbital and the hopping integral\\ in diatomic molecule cations subjected to harmonic excitation}
\author{I. A. Domagalska$^{\left(1\right)}$}
\author{M. W. Jarosik$^{\left(2\right)}$} 
\author{A. P. Durajski$^{\left(2\right)}$}
\author{J. K. Kalaga$^{\left(1,3\right)}$}
\author{R. Szcz{\c{e}}{\'s}niak$^{\left(2,4\right)}$}
\affiliation{$^1$ Quantum Optics and Engineering Division, Faculty of Physics and Astronomy, University of Zielona G{\'o}ra, 
                  Prof. Z. Szafrana 4a, 65-516 Zielona G{\'o}ra, Poland}
\affiliation{$^2$ Division of Physics, Cz{\c{e}}stochowa University of Technology, Ave. Armii Krajowej 19, 42-200 Cz{\c{e}}stochowa, Poland}
\affiliation{$^3$ Joint Laboratory of Optics of Palack\'{y} University and Institute of Physics of CAS, Faculty of Science, Palack\'{y} University, 17. listopadu 12, 771 46 Olomouc, Czech Republic} 
\affiliation{$^4$ Division of Theoretical Physics, Jan D{\l}ugosz University in Cz{\c{e}}stochowa, Ave. Armii Krajowej 13/15, 42-200 Cz{\c{e}}stochowa, Poland}
\date{\today}
\begin{abstract}
We analysed the dynamics of the positively charged ions of diatomic molecules (${\rm X_{2}^{+}}$ and ${\rm XY^{+}}$), in which the bond is realised 
by the single electron. We assumed that the atomic cores separated by the distance $R$ were subjected to the external excitation of the harmonic type with the amplitude $A$ and frequency $\Omega$. We found the ground states of ions using the variational approach within the formalism of second quantization (the Wannier function was reproduced by means of Gaussian orbitals). It occurred that, on the account of the highly non-linear dependence of the total energy on $R$, the chaotic dynamics of cores induced the chaotic evolution of the electronic Hamiltonian parameters (i.e. the energy of the electron orbital $\varepsilon$ and the hopping integral $t$). Changes in cation masses or in the charge arrangement does not affect qualitatively the values of Lyapunov exponents in the $A$-$\Omega$ parameter space. 
\end{abstract}
\maketitle

%%%%%%%%%%%%%%%%%%%%%%%%%%%%%%%%%%%%%%%%%%%%%%%%%%%%%%%%%%%%%%%%%%%%%%%%%%%%%%%%%%%%%%%%%%%%%%%%%%%%%%%%%%%%%%%%%%%%%%%%%%%%%%%%%%%%%%%%%%%%%%%%%%%%%%%%(I)
\section{Introduction}

While analysing the results of the conventional deterministic chaos theory, one can notice that the complicated dynamics of the physical system does not necessarily results from its intricate structure. Much more significance should be attached to the presence of the nonlinear interactions in the examined system, because they can lead to the exponential divergence of the initially close trajectories in the phase space. Let us recall the example of the 
three-body system with gravity interactions, which was examined in detail e.g. by Poincare \cite{Poincare1892A} (see~also~\cite{Garrido1982A}). Similar simple system (Lorenz equations) was succesfully applied also to the phenomenon of thermal convection \cite{Lorenz1963A}, or to the dynamics of the Belousov-Zhabotinsky chemical reaction \cite{Epstein1983A}. And the extremely simple system of this kind is the periodically accelerated pendulum for which the gravity force component implying the motion is proportional to the sine of the swing angle \cite{Collet1980A}. 

The conventional chaos theory is fairly well established by now and its results are presented in many scientific treatises \cite{Schuster1984A, Lin1984A, Cvitanovic1984A}. On the other hand, for the case of quantum systems, which are described in much more sophisticated way than the conventional ones, we cannot speak of full understanding of their dynamics yet 
\cite{Einstein1917A, Berry1983A, Casati1982B, Zaslavsky1981A, Haake2018A}. 
The researches carried out so far seem to suggest that there is no quantum system which would behave in the chaotic way 
(i.e. the one exhibiting the continuous power spectrum or deterministic diffusion) \cite{Schuster1984A}. This fact can be checked by studying the example of the Arnold's quantum transformation \cite{Hannay1980A} or stricken quantum rotator \cite{Casati1977A}. However, these quantum systems, which on approach to the boundary of validity of the conventional theory exhibit chaotic behaviour, have their wavefunctions distinctly different from the systems with regular behaviour at the same boundary. The wavefunctions of the free particle in the stadium and in the circle can be compared as the example \cite{McDonald1979A}. The reason for suppression of chaos in quantum system is said to be the finite value of the Planck constant ($h$), which along with the Heisenberg uncertainty principle introduces the indistinguishability of points in the \mbox{$2N$-dimensional} phase space contained within 
the $(h/2\pi)^{N}$ volume.

Currently developed research directions related to quantum chaos are based on the methods for solving quantum problems, where the perturbation cannot be considered small \cite{Courtney1995A, Courtney1995B, L96b, Stockmann2006A, Haake2018A}. In particular, the statistical descriptions of energy levels are used \cite{Courtney1996A, Heusler2007A, Mitchell2010A, Grass2013A, Frisch2014A, Riser2017A}. 
The starting point for considerations is the distribution of level spacing between eigenlevels:
$P\left(s\right)=\left<\delta\left(p-E_{j}+E_{j+1}\right)\right>$. For regular systems, $P\left(s\right)$ has the universal form of the Poisson distribution \cite{Berry1977A}: $P\left(s\right)=e^{-s}$, i.e. the successive energy levels are not correlated. 
The universal nature of the distribution means that it is valid for systems belonging to the same symmetry class and does not depend on their individual properties. Based on the Random Matrix Theory, it was shown that in the case of quantum chaotic systems three basic universal distributions can be distinguished: the Gaussian Orthogonal Ensemble, the Gaussian Unitary Ensemble, and the Gaussian Symplectic Ensemble \cite{Dyson1962A, Porter1965A, Bohigas1984A, Guhr1998A}. Other chaotic criteria were also obtained, e.g. the spectral stiffness \cite{Dyson1963A}, 
the autocorrelation function of energy levels velocity \cite{Simons1993A, Schafer1993A}, and the noise of $1/f$ type \cite{Relano2002A}.  
Please note that the $1/f$ feature is universal, i.e. this behavior is the same for all kinds of chaotic systems, independently of their symmetries. Another approach is based on the semiclassical methods such as periodic-orbit theory connecting the classical trajectories of the dynamical system 
with the quantum features \cite{Vogl2017A}. In addition, studies that directly refer to the correspondence principle are worth emphasizing \cite{Stockmann2006A, Haake2018A}.
At this point, it is worth noting that also other methods were implemented in quantum chaos' detection. For instance, the fidelities of the wave-functions (or the density matrices)  have been applied for such purposes \cite{WLT02, EWL02, KKL09}. Also, the parameter based on nonclassicality criterion \cite{KZ04} has been proposed as a that which can be used in the detection of quantum chaotic behavior \cite{KKL08}. Finally, Kullback-{L}eibler quantum divergence has also been proposed as an indicator of the quantum chaotic characteristic of the nonlinear system appears. \cite{KKL12}.

The presented paper analyses the dynamics of positively charged ions of diatomic molecules (${\rm X_{2}^{+}}$ or ${\rm XY^{+}}$), which atomic cores are subjected to the harmonic excitation. The considered systems are rather unique from the chaos teory point of view, because they consist of two correlated subsystems of very simple structure: the conventional one (the atomic cores or - in the extreme case of hydrogen - protons) and the purely quantum 
one (the electron). Let us notice that this description \mbox{of the molecule} is based on the Born-Oppenheimer approximation \cite{Born1927A}, which makes use of the fact that the atomic cores can have thousands of times greater mass than the single electron. Therefore their motion is slower by several orders of magnitude than the motion of electrons. Reversely, electrons adapt themselves 'immediately' to the changed position of cores. Because of this one can examine the influence of the chaotic dynamics of atomic cores, resulting from the existence of the highly nonlinear internuclear potential, 
on the time evolution of the parameters of the electronic Hamiltonian (energy of the electron orbital $\varepsilon$ and the hopping integral $t$). This does not mean that the quantum subsystem (the electronic one) will evolve in the chaotic manner. Nevertheless it will respond to the behaviour of atomic cores, so that the model described here can serve as the basis for investigation of the changing dynamics of electrons in order to determine the influence of the chaotic evolution of the core subsystem on the quantum electronic system.
   
The structural simplicity of the considered systems is well worth attention, since it plays the considerable role. It enables to perform complicated quantum-mechanic calculations with the utmost accuracy (which is demanded in quantum chemistry) \cite{Born1927A, Kolos1964A, Kolos1968A, Jarosik2018A}. We performed calculations for the presented work with an accuracy to six decimal places.

It is worth noticing that during the performed analysis we took into account cations with different core masses and the asymmetric charge distribution between the cores. It allowed us to show the universal character of the chaotic behaviour of cores in the whole family of diatomic cations with the molecular bond realised by the single electron.

%%%%%%%%%%%%%%%%%%%%%%%%%%%%%%%%%%%%%%%%%%%%%%%%%%%%%%%%%%%%%%%%%%%%%%%%%%%%%%%%%%%%%%%%%%%%%%%%%%%%%%%%%%%%%%%%%%%%%%%%%%%%%%%%%%%%%%%%%%%%%%%%%%%%%%%(II)
\section{Description of the ground state of cations with one-electron bond}

Let us consider ions composed of two either identical (X) or different (X and Y) atoms, while we regard their atomic nuclei with the inner shell electrons as the atomic cores. The effective charges of these cores are $Z_{1}$ and $Z_{2}$, where $Z_{1}+Z_{2}=+2$ (in atomic units). The atomic cores are bound to form the molecule by means of the single electron. In the considered case either the ${\rm X_{2}^{+}}$ or the ${\rm XY^{+}}$ cation arises. 
The simplest example of such the system is the ${\rm H^{+}_{2}}$ or ${\rm D^{+}_{2}}$ molecule \cite{Schaad1970A}, however the core charge asymmetry does not occur there ($Z_{1}=Z_{2}=+1$). But the above presented description can also be applied to the more complicated systems, e.g. the homo and heteronuclear alkali-metal cation dimers 
($\rm Li_{2}^{+}$, $\rm Na_{2}^{+}$, $\rm LiNa^{+}$, $\rm K_{2}^{+}$, and $\rm LiH^{+}$); the cation $\rm {Cu_{2}^{+}}$ or other examples
\cite{Claxton1974A, Hudson1977A, Shida1981A, Hoefelmeyer2000A, Klusik1981A, Chiu2007A, Cataldo2001A, Pilon2010A, Moret2013A, Sousa2017A}. 
It is also worth noting that the system consisting of two nuclei and one electron was already studied in the twenties of the last century
\cite{Alexandrow1926A, Wilson1928A}.

Let us take into account the total energy of an exemplary diatomic cation: 
$E_{T}=E_{c}+E_{e}$, where $E_{c}=2Z_{1}Z_{2}/R$, represents the energy of the core-core interaction ($R$ stands for the intercore 
distance $R=|{\bf R}|$). The energy of the electron orbital in the ground state is denoted by $E_{e}$ symbol. It can be calculated with the use of Hubbard Hamiltonian written in the second quantization notation \cite{Hubbard1963A, Hubbard1964A, Fetter1971A, Spalek2015A}:  
\begin{eqnarray}
\label{r01-II}
\hat{\mathcal{H}}_{e}&=&\varepsilon_{1}\hat{n}_1+\varepsilon_{2}\hat{n}_2
       +\sum_{\sigma}\left(t_{12}\hat{n}_{12\sigma}+t_{21}\hat{n}_{21\sigma}\right), 
\end{eqnarray}
where: $\hat{n}_{j}=\sum_{\sigma}\hat{n}_{j\sigma}$, $\hat{n}_{j\sigma}=\hat{c}^{\dag}_{j\sigma}\hat{c}_{j\sigma}$, and 
$\hat{n}_{ij\sigma}=\hat{c}^{\dag}_{i\sigma}\hat{c}_{j\sigma}$. The symbol $\hat{c}^{\dag}_{j\sigma}$ ($\hat{c}_{j\sigma}$) represents the creation (anihilation) operator referring to the electronic state of the spin $\sigma\in\left\{\uparrow,\downarrow\right\}$ on the $j^{th}$ core. The energetic parameters of the Hamiltonian should be calculated numerically directly from their definitions \cite{Acquarone1998A}:
\begin{eqnarray}
\label{r02-II}
\varepsilon_{i} &=&
\int d^{3}{\bf r}\Phi_{i}\left({\bf r}\right)\left[-\nabla^{2}-\frac{2Z_{i}}{|{\bf r}-{\bf R}|}\right]\Phi_{i}\left({\bf r}\right),\\
t_{ij} &=&
\int d^{3}{\bf r}\Phi_{i}\left({\bf r}\right)\left[-\nabla^{2}-\frac{2Z_{j}}{|{\bf r}-{\bf R}|}\right]\Phi_{j}\left({\bf r}\right).
\end{eqnarray}
The symbol $\Phi_{j}\left({\bf r}\right)$ denotes the Wannier function: 
\begin{eqnarray}
\label{r03-II}
\Phi_{1}\left({\bf r}\right)&=&
A_{+}\left(S\right)\phi_{1}\left({\bf r}\right)+
A_{-}\left(S\right)\phi_{2}\left({\bf r}\right),\\ \nonumber
\Phi_{2}\left({\bf r}\right)&=&
A_{-}\left(S\right)\phi_{1}\left({\bf r}\right)+
A_{+}\left(S\right)\phi_{2}\left({\bf r}\right),
\end{eqnarray}
where the normalisation constants take the form:
\begin{eqnarray}
\label{r04-II}
A_{\pm}\left(S\right)=\frac{1}{2}\left[\frac{1}{\sqrt{1+S}}\pm
\frac{1}{\sqrt{1-S}}\right].
\end{eqnarray}
The overlap integral ($S$) can be determined from the formula: 
$S=\int d^{3}{\bf r}\phi_{1}\left({\bf r}\right)\phi_{2}\left({\bf r}\right)=\exp\left(-\alpha^{2}R^{2}/2\right)$, where the Gaussian orbital is given by the expression: $\phi_{i}\left({\bf r}\right)=\left(2\alpha^{2}/\pi\right)^{3/4}\exp\left[-\alpha^{2}({\bf r}-{\bf R}_{i})^{2}\right]$, and $\alpha$ is the variational parameter. It is worth noticing that the the Hamiltonian~\eq{r01-II}, despite its simplicity, takes into account all contributions to the energy of the electron orbital due to the fact that we consider cations with the one-electron bond. In the case of the multiple bond, the energies of electronic correlations should be additionally included. For the simplest case of the double bond they comprise the on-site Coulomb repulsion $U$, the energy of the inter-site Coulomb repulsion $K$, the exchange integral $J$, and the energy of correlated electron hopping $V$. These quantities are thoroughly discussed e.g. in \cite{Spalek2015A}.  

The electronic Hamiltonian in matrix notation takes the form:
\begin{equation}
\label{r05-II}
\hat{\mathcal{H}}_{e }=\left(
\begin{array}{cccc}
\varepsilon_{1}             &  0                        & t_{12}                     & 0                   \\
0                           &  \varepsilon_{1}          & 0                          & t_{12}              \\
t_{21}                      &  0                        & \varepsilon_{2}            & 0                   \\
0                           &  t_{21}                   & 0                          & \varepsilon_{2}           
\end{array}
\right),
\end{equation}
for the basis assumed as follows:
\begin{eqnarray}
\label{r06-II}
|1_{A}, 1\slash 2>&=&|(1,0),(0,0)>=\hat{c}_{1\uparrow}^{\dag}|0>,\\
|1_{A},-1\slash 2>&=&|(0,1),(0,0)>=\hat{c}_{1\downarrow}^{\dag}|0>,\\
|1_{B}, 1\slash 2>&=&|(0,0),(1,0)>=\hat{c}_{2\uparrow}^{\dag}|0>,\\
|1_{B},-1\slash 2>&=&|(0,0),(0,1)>=\hat{c}_{2\downarrow}^{\dag}|0>.
\end{eqnarray}
We introduced the notation $|w_{x},s>$, where $w$ denotes the maximum number of electrons at the site $A$ ($x=A$) or $B$ ($x=B$). The symbol $s$ represents the resultant spin ($s\in\{-1\slash 2,1\slash 2\}$). The notation $|0>=|(0,0),(0,0)>$ describes the vacuum state. It should be clearly stated that the second quantization formalism is absolutely equivalent to the formalism of wave mechanics introduced by Schr{\"o}dinger \cite{Schrodinger1926A, Schrodinger1926B, Schrodinger1926C, Schrodinger1926D}.

The eigenvalues of the $\hat{\mathcal{H}}_{e}$ Hamiltonian can be calculated analitycally:
\begin{eqnarray}
\label{r07-II}
E_{1}=E_{2}&=&\frac{\varepsilon_{1}+\varepsilon_{2}}{2}-\sqrt{t_{12}t_{21}+\left(\frac{\varepsilon_{1}-\varepsilon_{2}}{2}\right)^{2}}\\ \nonumber
&=&E_{min},
\end{eqnarray}
\begin{eqnarray}
\label{r09-II}
E_{3}=E_{4}&=&\frac{\varepsilon_{1}+\varepsilon_{2}}{2}+\sqrt{t_{12}t_{21}+\left(\frac{\varepsilon_{1}-\varepsilon_{2}}{2}\right)^{2}}.
\end{eqnarray}
It can be easily seen that the ground state is the degenerated one, this being related to the existence of two directions of the electronic spin projections. The degeneration can be removed by the constant external magnetic field ($H$) applied to the cations: 
$\hat{\mathcal{H}}_{H}=-2H\left(\hat{S}^{z}_{1}+\hat{S}^{z}_{2}\right)$, where 
$\hat{S}^{z}_{j}=\frac{1}{2}\left(\hat{n}_{j\uparrow}-\hat{n}_{j\downarrow}\right)$ \cite{Drzazga2019A, Schiff1949A}. The eigenvectors have the form:
\begin{eqnarray}
\label{r11-II}
|1,1>=e_{2}\left(\frac{1}{e_{3}}\hat{c}^{\dag}_{1\uparrow}+\frac{e^{-}_{1}}{e_{4}}\hat{c}^{\dag}_{2\uparrow}\right)|0>,
\end{eqnarray}
\begin{eqnarray}
\label{r12-II}
|2,1>=e_{2}\left(\frac{1}{e_{3}}\hat{c}^{\dag}_{1\downarrow}+\frac{e^{-}_{1}}{e_{4}}\hat{c}^{\dag}_{2\downarrow}\right)|0>,
\end{eqnarray}
\begin{eqnarray}
\label{r13-II}
|3,1>=-e_{2}\left(\frac{1}{e_{3}}\hat{c}^{\dag}_{1\uparrow}+\frac{e^{+}_{1}}{e_{4}}\hat{c}^{\dag}_{2\uparrow}\right)|0>,
\end{eqnarray}
\begin{eqnarray}
\label{r14-II}
|4,1>=-e_{2}\left(\frac{1}{e_{3}}\hat{c}^{\dag}_{1\downarrow}+\frac{e^{+}_{1}}{e_{4}}\hat{c}^{\dag}_{2\downarrow}\right)|0>.
\end{eqnarray}
We introduced some auxiliary designations in the above formulae, namely:
\begin{eqnarray}
\label{r15-II}
e^{\pm}_{1}&=&\varepsilon_{2}-\varepsilon_{1}\pm\sqrt{4t_{12}t_{21}+\left(\varepsilon_{1}-\varepsilon_{2}\right)^2},\\
e_{2}&=&\sqrt{2t_{12}\left(t_{12}+t_{21}\right)-\left(\varepsilon_{1}-\varepsilon_{2}\right)e^{+}_{1}},\\
e_{3}&=&\sqrt{2}\sqrt{\left(t_{12}+t_{21}\right)^2+\left(\varepsilon_{1}-\varepsilon_{2}\right)^2},\\
e_{4}&=&2t_{12}e_{3}.
\end{eqnarray}
\begin{figure}         
\includegraphics[width=\columnwidth]{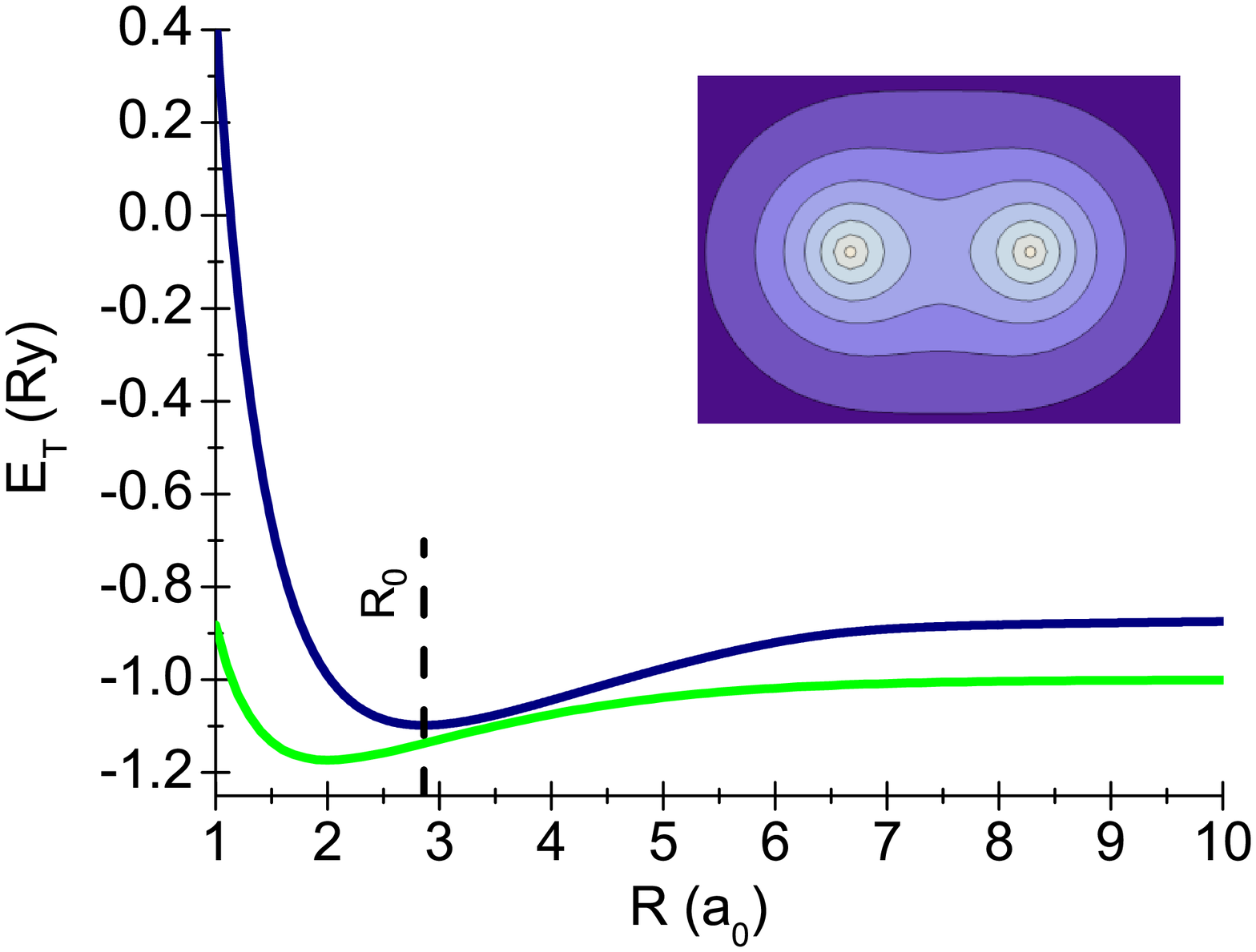}
\caption{
         Total energy ($E_{T}$) of the ion of molecule $\rm H^{+}_{2}$ versus the intercore distance $R$ (the Gaussian orbitals).
         The inset presents the distribution of the electronic charge in the equilibrium state ($R=R_{0}$).
         The green line indicates the results obtained for $\rm H^{+}_{2}$ ion by using the Slater-type orbitals. 
         }
\label{f01II}
\includegraphics[width=\columnwidth]{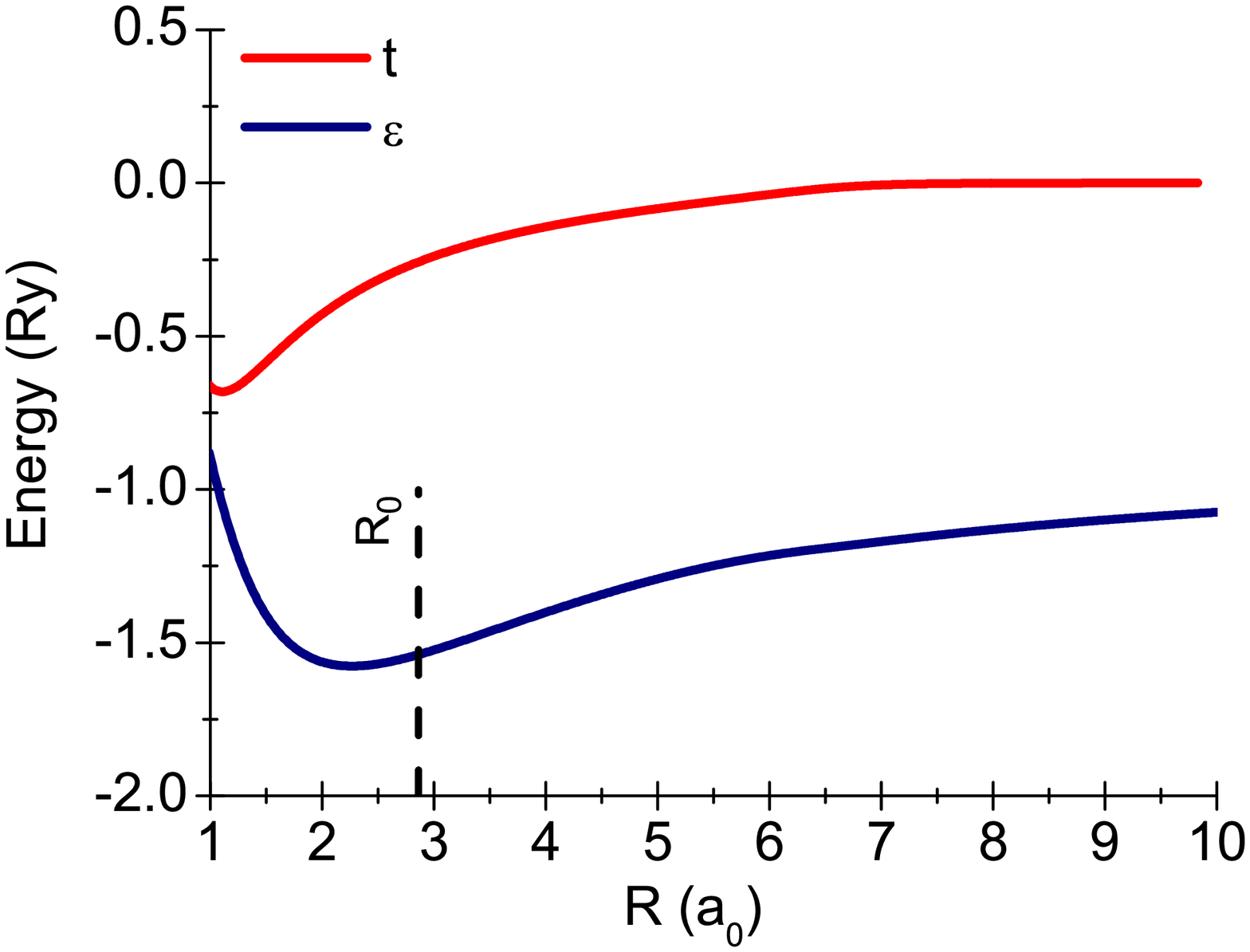}
\caption{
         The dependence of the energy of the electron orbital $\varepsilon$ and the hopping integral $t$ on the intercore distance $R$ 
         for the case of $\rm H^{+}_{2}$ ion. The results were obtained using the Gaussian orbitals.
         }
\label{f02II}
\end{figure}
\begin{figure}
\includegraphics[width=\columnwidth]{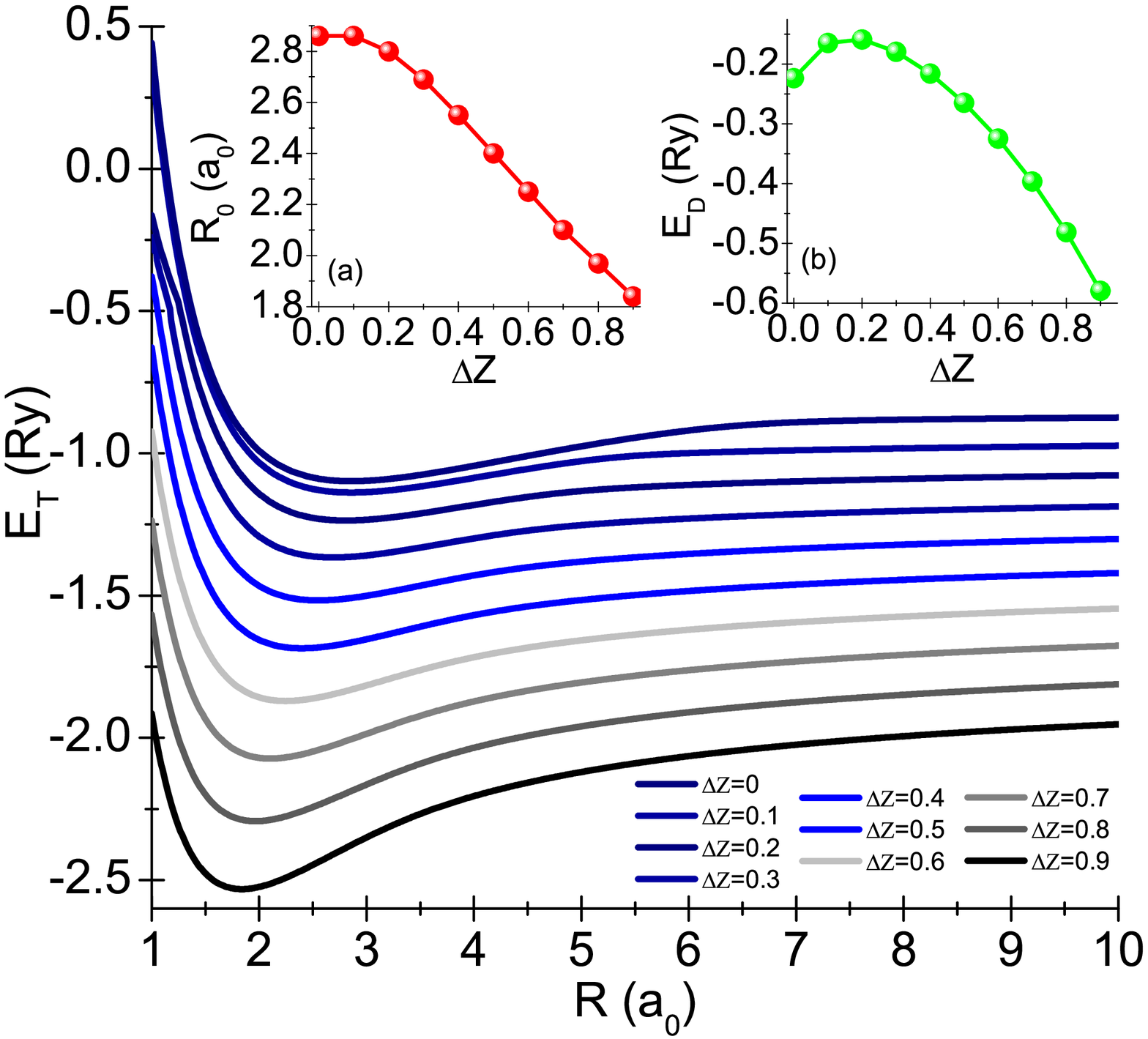}
\caption{
         The influence of the charge asymmetry of the atomic cores on the energetic state of the ${\rm XY^{+}}$ ion. 
         The insets show the dependence of $R_{0}$ and $E_{D}$ on $\Delta Z$. 
         The value $\Delta Z=0.3$ corresponds to the cation $\rm LiH^{+}$ and 
                          the value $\Delta Z=0.9$ corresponds to the $\rm LiNa^{+}$ system.}  
\label{f03II}
\includegraphics[width=\columnwidth]{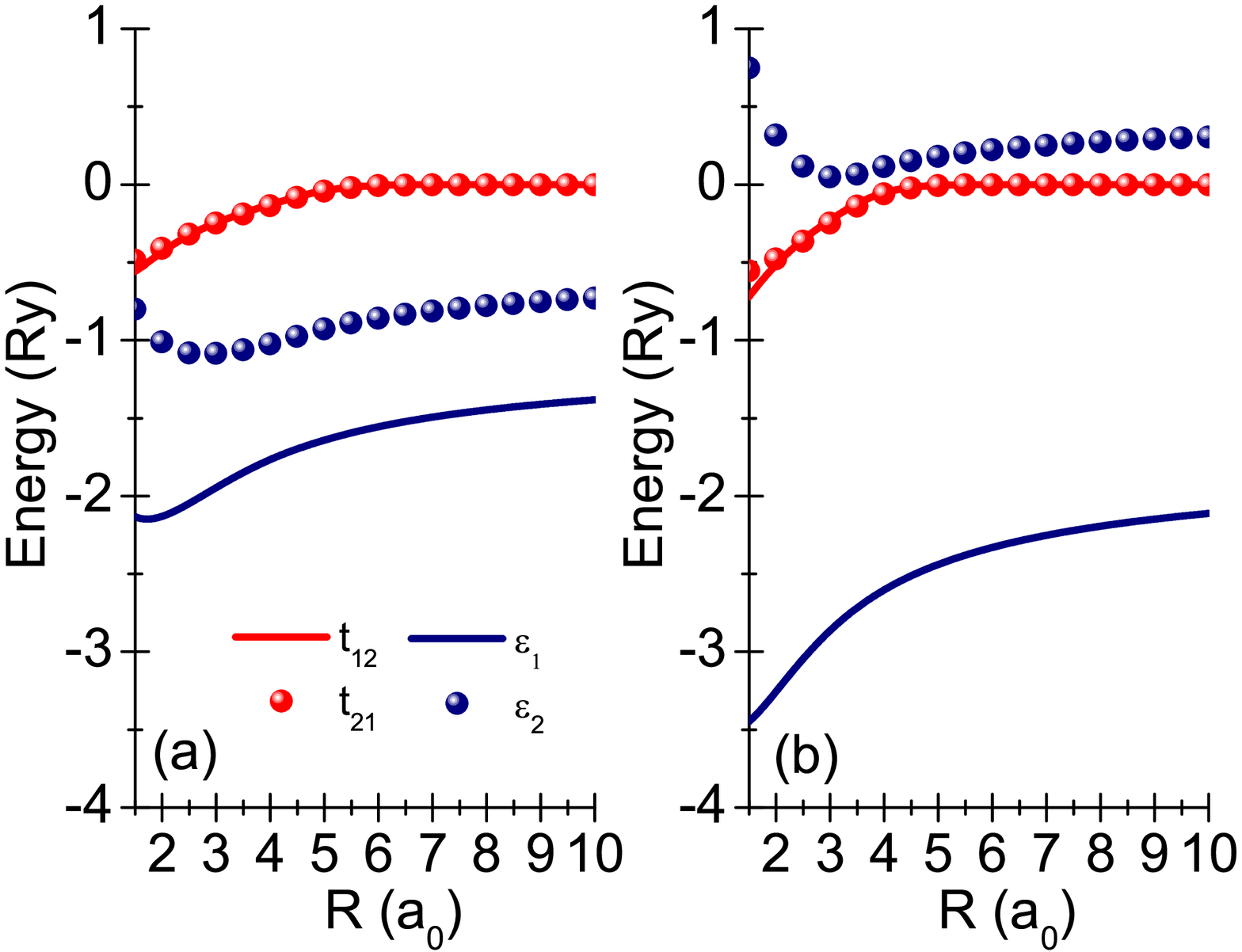}
\caption{
         Energies of the electron orbital $\varepsilon_{1}$ and $\varepsilon_{2}$, as well as the hopping integrals $t_{12}$ and $t_{21}$ versus 
         the intercore distance $R$. 
         (a) The case of the $\rm LiH^{+}$ cation ($\Delta Z=0.3$). (b) The case of the $\rm LiNa^{+}$ cation ($\Delta Z=0.9$).
         }  
\label{f04II}     
\end{figure}
\begin{figure}         
\includegraphics[width=\columnwidth]{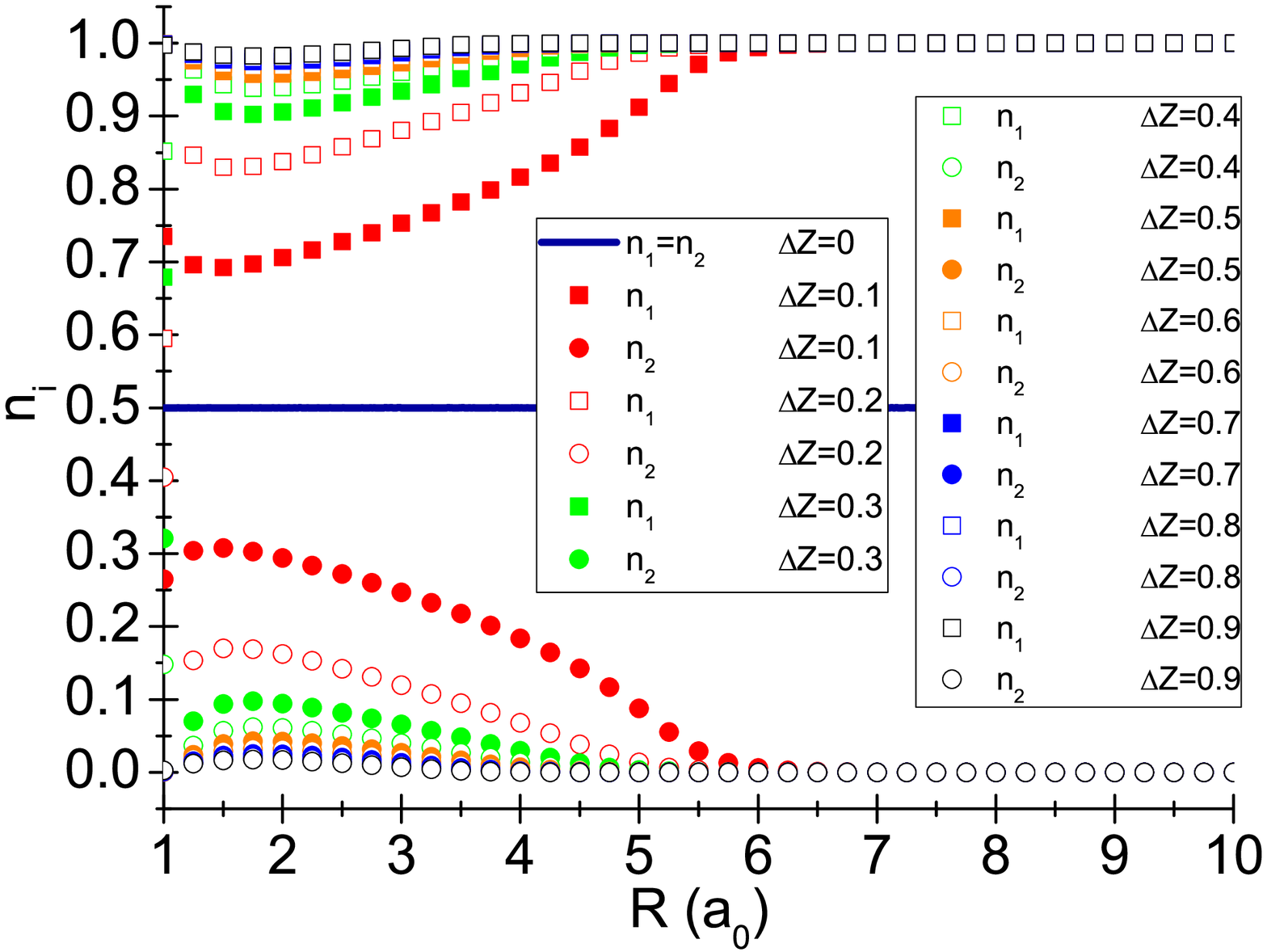}
\caption{
         Occupation at the ${\rm XY^{+}}$ ion sites for the increasing charge asymmetry of the atomic cores. 
        The value $\Delta Z=0.3$ describes the system $\rm LiH^{+}$ and 
                          the value $\Delta Z=0.9$ corresponds to the $\rm LiNa^{+}$ system.}          
\label{f05II}
\end{figure}

The first step of the analysis focuses on the case of charge symmetry ($Z_{1}=Z_{2}=+1$). 

Figure \fig{f01II} presents the total energy of ${\rm H^{+}_{2}}$ ion versus the intercore distance $R$ 
(again for the case of the Wannier function comprises the $1s$ Slater-type orbitals \cite{Kadzielawa2014A, Jarosik2018A}). 
The $E_{T}\left(R\right)$ function exhibit the characteristic mimimum at $R_{0}=2.85818$~${\rm a_{0}}$ 
($a_{0}\simeq 0.529\cdot 10^{- 10}$~m) of the value equal to $E_{0}=E_{T}\left(R_{0}\right)=-1.09797$~Ry. 
From the physical viewpoint $R_{0}$ determines the equilibrium distance of the system, which corresponds to the dissociation energy 
$E_{D}=E_{0}-\lim_{R\rightarrow+\infty}E_{T}\left(R\right)=-0.24658$~Ry. The background in Figure \fig{f01II} shows the equilibrium distribution of the electronic charge for the ${\rm H^{+}_{2}}$ ion ($\rho\left({\bf r}\right)=\sum_{j}\Phi^{\star}_{j}\left({\bf r}\right)\Phi_{j}\left({\bf r}\right)$).
For the case of Slater-type orbitals (see also Figure \fig{f01II}), we came to the following estimates: 
$R_{0}=2.00330$~${\rm a_{0}}$, $E_{0}=-1.17301$~Ry, and $E_{D}=-0.17241$~Ry. These results agree with the results of the numerical analysis carried out 
by Schaad and Hicks: $R_{0}=1.9972$~${\rm a_{0}}$ and $E_{0}=-1.20527$~Ry \cite{Schaad1970A}.  

Using the value of $R_{0}$ for ${\rm H^{+}_{2}}$ ion (the Gaussian orbitals), we calculated equilibruim values of the energy of the electron orbital $\varepsilon_{0}$ and the hopping integral $t_{0}$. The results are: $-1.53949$~Ry and $-0.25823$~Ry, respectively. The value of the variational parameter $\alpha_{0}$ is equal to $0.66410$~${\rm a^{-1}_{0}}$. 
For the case of Slater-type orbitals, we obtained: $\varepsilon_{0}=-1.69825$~Ry, $t_{0}=-0.47312$~Ry, and 
$\alpha_{0}=1.23803$~${\rm a^{-1}_{0}}$. The full dependence of the energy of the electron orbital and the hopping integral on the intercore distance for the $\rm H^{+}_{2}$ cation is presented in Figure \fig{f02II}. It can be seen that the energetic parameters of the Hamiltonian depend strongly on $R$. 

\begin{table*}
\caption{
The electronic parameters and the occupation at site $j$ for selected values of $\Delta Z$.
The stable ions $\rm LiH^{+}$ and $\rm LiNa^{+}$ are characterized by the following values: 
                 $\Delta Z=0.3$ and $\Delta Z=0.9$. 
}
\begin{tabular}{c|cccccccc} 
      &     &          &           &           &           &           &          &          \\
$\Delta Z$ & $ $ & $\varepsilon_{1_{0}}$~$\rm{\left[Ry\right]}$  & $\varepsilon_{2_{0}}$~$\rm{\left[Ry\right]}$ & $t_{12_{0}}$~$\rm{\left[Ry\right]}$ & $t_{21_{0}}$~$\rm{\left[Ry\right]}$ & $n_{1}$ & $n_{2}$ & $n_{12}=n_{21}$ \\
\hline \\
0     & $ $ & -1.5393  & -1.5393   & -0.257969 & -0.257969 & 0.5       & 0.5       & 0.5      \\
0.1   & $ $ & -1.68416 & -1.39315  & -0.259146 & -0.258067 & 0.745972  & 0.254028  & 0.435313 \\
0.2   & $ $ & -1.83909 & -1.24525  & -0.270709 & -0.268424 & 0.87115   & 0.12885   & 0.335034 \\
{\bf 0.3} & $ $ & {\bf -2.00904} & {\bf -1.08618}  & {\bf -0.292397} & {\bf -0.288287} & {\bf 0.924195}  & {\bf 0.0758051} & {\bf 0.264686} \\
0.4   & $ $ & -2.19494 & -0.908019 & -0.322854 & -0.31562  & 0.949019  & 0.0509811 & 0.219959 \\
0.5   & $ $ & -2.39521 & -0.70532  & -0.360416 & -0.348134 & 0.962394  & 0.0376063 & 0.190242 \\
0.6   & $ $ & -2.60878 & -0.474263 & -0.404472 & -0.384519 & 0.970489  & 0.0295114 & 0.169235 \\
0.7   & $ $ & -2.83662 & -0.209848 & -0.456342 & -0.424812 & 0.975819  & 0.024181  & 0.155278 \\
0.8   & $ $ & -3.07207 & 0.0834013 & -0.51086  & -0.464853 & 0.979696  & 0.020304  & 0.141038 \\
{\bf 0.9} & $ $ & {\bf -3.32195} & {\bf 0.416627} & {\bf -0.574607} & {\bf -0.507966} & {\bf 0.982581} & {\bf 0.017419} & {\bf 0.130827} \\
\end{tabular}
\label{t01}
\end{table*} 

Let the symbol $\hat{O}$ represent the operator corresponding to the given physical quantity. We determine the observables by means of the formula:
$\left<\hat{O}\right>=\left<w|\hat{O}|w\right>$, where $|w>$ denotes the eigenvector of the electronic Hamiltonian corresponding to the minimum value of the total energy \cite{Schiff1949A}. As far as we consider the positively charged ion of the molecule with one-electron bond, only one observable is physically interesting - the occupation at site~$j$: 
$n_{j}=\left<\hat{n}_{j}\right>=\sum_{\sigma}\left<\hat{n}_{j\sigma}\right>$. Other observables either are the re-scaled $n_{j}$ values or take the zero value \cite{Acquarone1998A}. If we choose an electron with the upward directed spin, we can calculate $n_{j}$ analytically from the formulae:
\begin{eqnarray}
\label{r16-II}
{n}_{1}&=&\left(\frac{e_{2}}{e_{3}}\right)^{2}, \\ \nonumber
{n}_{2}&=&\left(\frac{e_{2}e^{-}_{1}}{e_{4}}\right)^{2}=1-{n}_{1},\\ \nonumber
{n}_{12}&=&\frac{e^{2}_{2}e^{-}_{1}}{e_{3}e_{4}}={n}_{21}.
\end{eqnarray}
The same results we would obtain also for the downward directed spin.

Let us discuss now the physical state of the ${\rm XY^{+}}$ cation, the case of asymmetric distribution of the core charge ($Z_{1}\neq Z_{2}$). Figure \fig{f03II} illustrates the influence of the increasing charge asymmetry of the atomic cores constituting the ${\rm XY^{+}}$ ion on the shape of the $E_{T}\left(R\right)$ function. One can notice that the equilibrium distance $R_{0}$ shortens with an increase in the parameter $\Delta Z=|Z_{1}-Z_{2}|$ (inset (a)). The dissociation energy $E_{D}$ at first slightly increases, then decreases (inset (b)). 
The equilibrium values of the electronic Hamiltonian parameters can be found in Table \tab{t01}. 

The results collected in Figure \fig{f03II} and in Table \tab{t01} can be referred to stable cations 
$\rm LiH^{+}$ and $\rm LiNa^{+}$ analyzed in the literature \cite{Sousa2017A}. In the first step, let's calculate the effective nuclear charge 
$Z=N-\sigma$ for H, Li, and Na atoms. The symbol $N$ denotes the atomic number and $\sigma$ is the Shielding or screening constant. 
The parameter $\sigma$ is the sum of the following contributions:
(i) Each other electron in the same group as the electron of interest shield to the extent of $0.35$ nuclear charge units - except $1s$ group, 
in which the other electron contributes only $0.30$.
(ii) If the group $n$ is of $s, p$ type, the amount of $0.85$ from each electron in $n-1$-{\it th} shell and the amount of $1$ for each 
electron from $n-2$ and lower shells is added to the shielding constant. 
(iii) If the group is of $d$ or $f$ type, the amount is equal to $1$ for each electron.

The electronic configurations H, Li, and Na atoms are in the form: 
$1s^{1}$, $1s^{2}2s^{1}$, and $1s^{2}2s^{2}2p^{6}3s^{1}$, respectively. From here, we get: 
$Z_{\rm H}=1$, $Z_{\rm Li}=1.3$, and $Z_{\rm Na}=2.2$. The calculations give: 
$\Delta Z_{\rm LiH^{+}}=0.3$ and $\Delta Z_{\rm LiNa^{+}}=0.9$.

Figure \fig{f04II} shows full trajectories of the energies of the electron orbital and the hopping integrals versus $R$ 
for $\rm LiH^{+}$. The obtained results prove that the charge asymmetry of the atomic cores can induce great differences in values 
of the considered quantities. It should be also noticed that the change in the shape of the $E_{T}\left(R\right)$ function induced by the asymmetric distribution of the charge distribution on the atomic cores will result in the noticeable change in phonon properties and the values of 
the electron-phonon coupling function~\cite{Wrona2019A}. 

The influence of the charge asymmetry of atomic cores on the occupation at site $j$ is represented in Figure \fig{f05II}. 
The selected values of $n_{1}$ and $n_{2}$ are gathered in Table \tab{t01}. We can observe the same probability of finding the electron on either core for the symmetric case $Z_{1}=Z_{2}$. As expected, the charge asymmetry increases the occupation at the atomic core with higher $Z_{j}$ value. 

%%%%%%%%%%%%%%%%%%%%%%%%%%%%%%%%%%%%%%%%%%%%%%%%%%%%%%%%%%%%%%%%%%%%%%%%%%%%%%%%%%%%%%%%%%%%%%%%%%%%%%%%%%%%%%%%%%%%%%%%%%%%%%%%%%%%%%%%%%%%%%%%%%%%%%(III)
\section{Time evolution of the energy of the electron orbital and the hopping integral}
\begin{figure*}         
\includegraphics[width=0.6\textwidth]{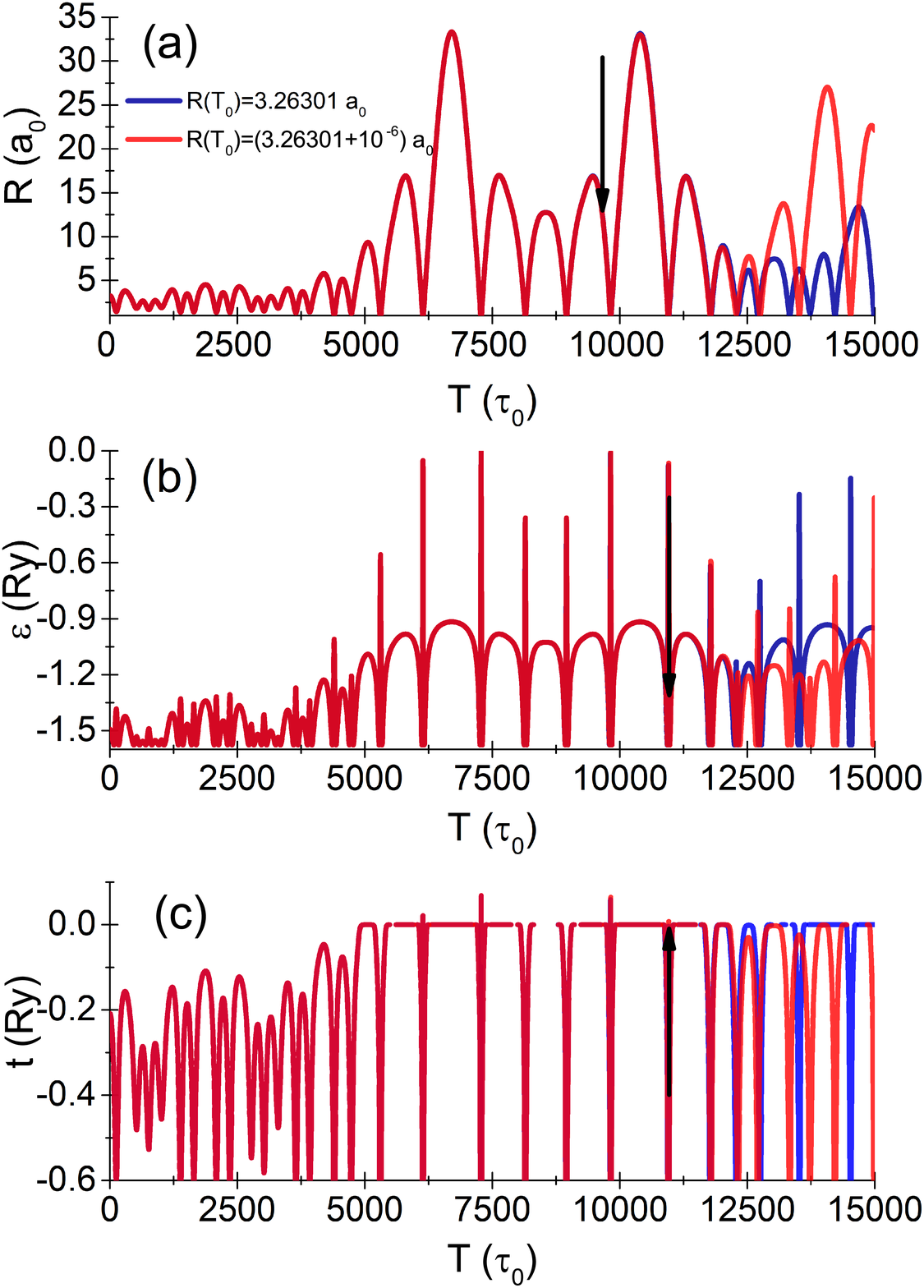}
\caption{
        (a) Time dependence of the intercore distance of ${\rm H^{+}_{2}}$ cation (the case of Gaussian orbitals) 
        for two very close trajectories $R_{1}\left(T\right)$ and $R_{2}\left(T\right)$, initially $10^{-6}$~${\rm a_{0}}$ apart. 
        The folowing parameters of the exciting force were assumed: $A=0.3$~${\rm a_{0}}$ and $\Omega=0.06$~${\rm \tau^{-1}_{0}}$. 
        (b) and (c) Chaotic evolution of the energy of the electron orbital and the hopping integral. Black arrows point to the values 
        of Lyapunov time.
        }        
\label{f01III}
\end{figure*}
\begin{figure*}         
\includegraphics[width=\textwidth]{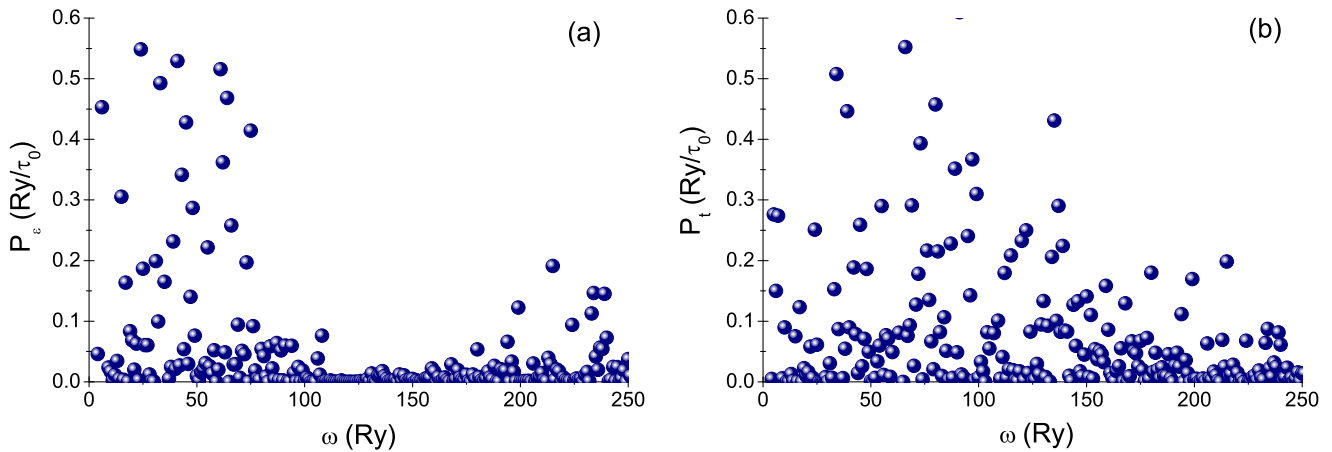}
\caption{Power spectra $P_{\varepsilon}\left(\omega\right)$ and $P_{t}\left(\omega\right)$ of the $\varepsilon(T)$ and the $t(T)$ functions.
         The case of ${\rm H^{+}_{2}}$ cation (the Gaussian orbitals).}
\label{f02III}
\end{figure*}
\begin{figure*}         
\includegraphics[width=0.32\linewidth]{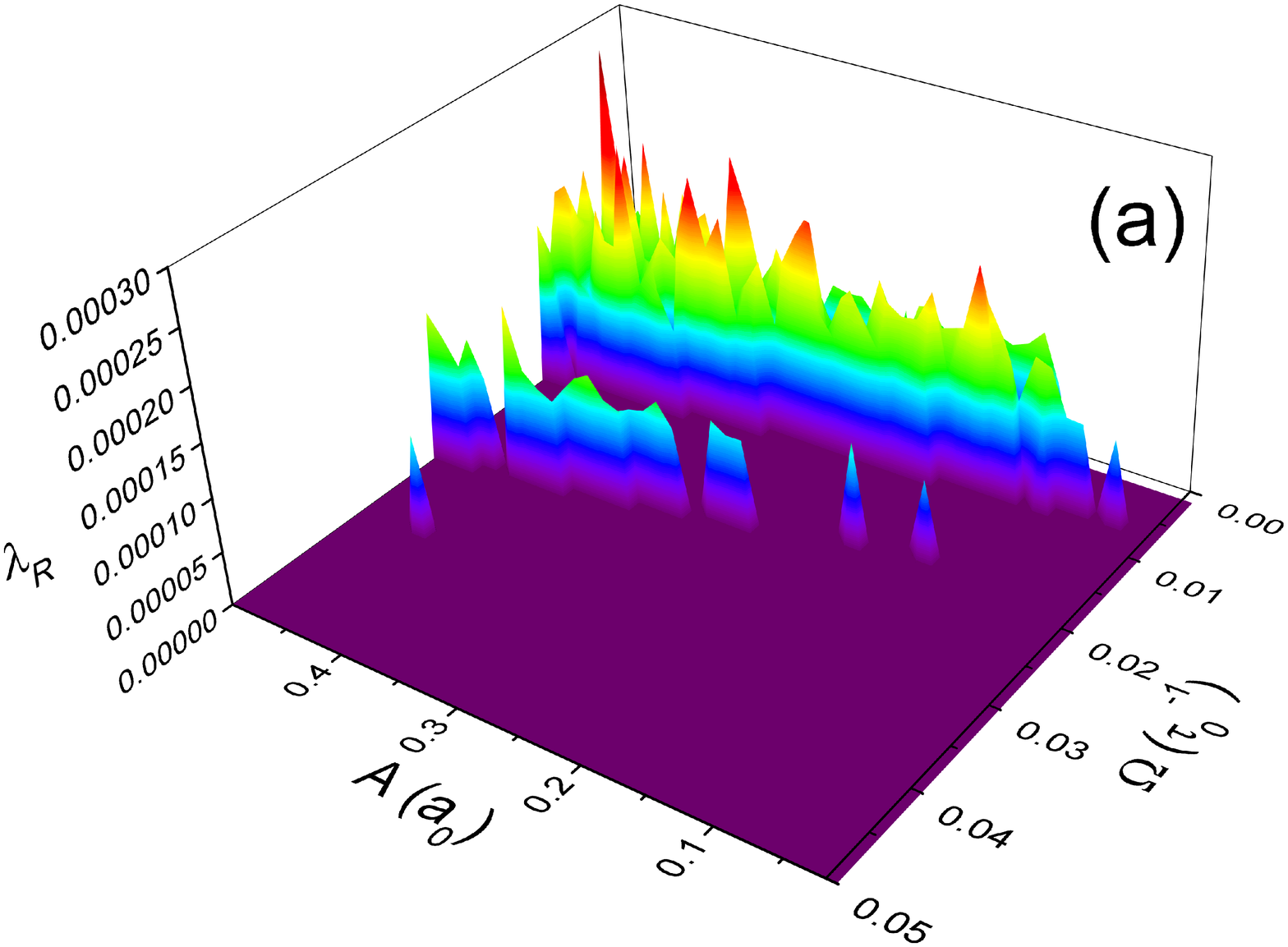}
\includegraphics[width=0.32\linewidth]{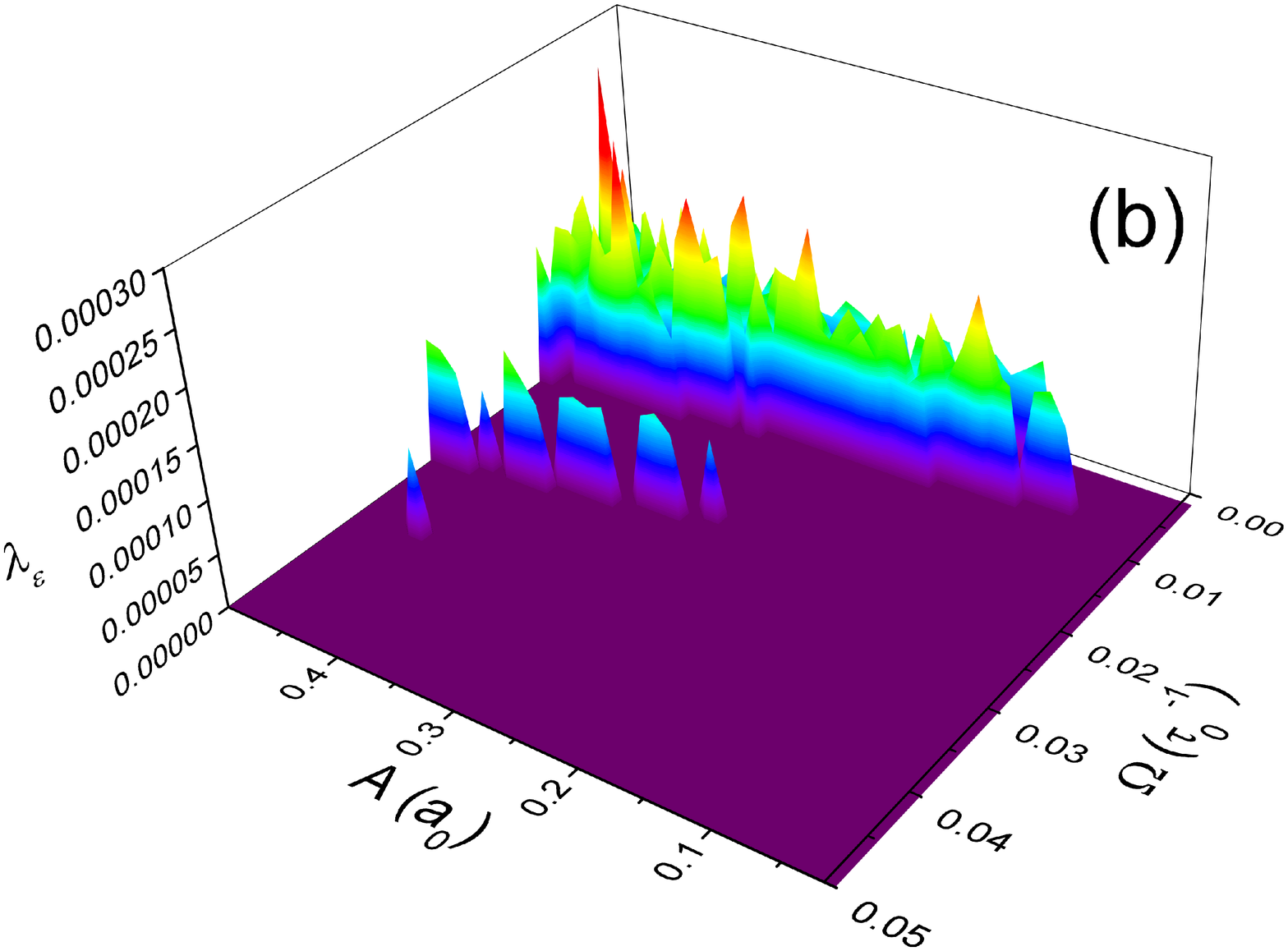}
\includegraphics[width=0.32\linewidth]{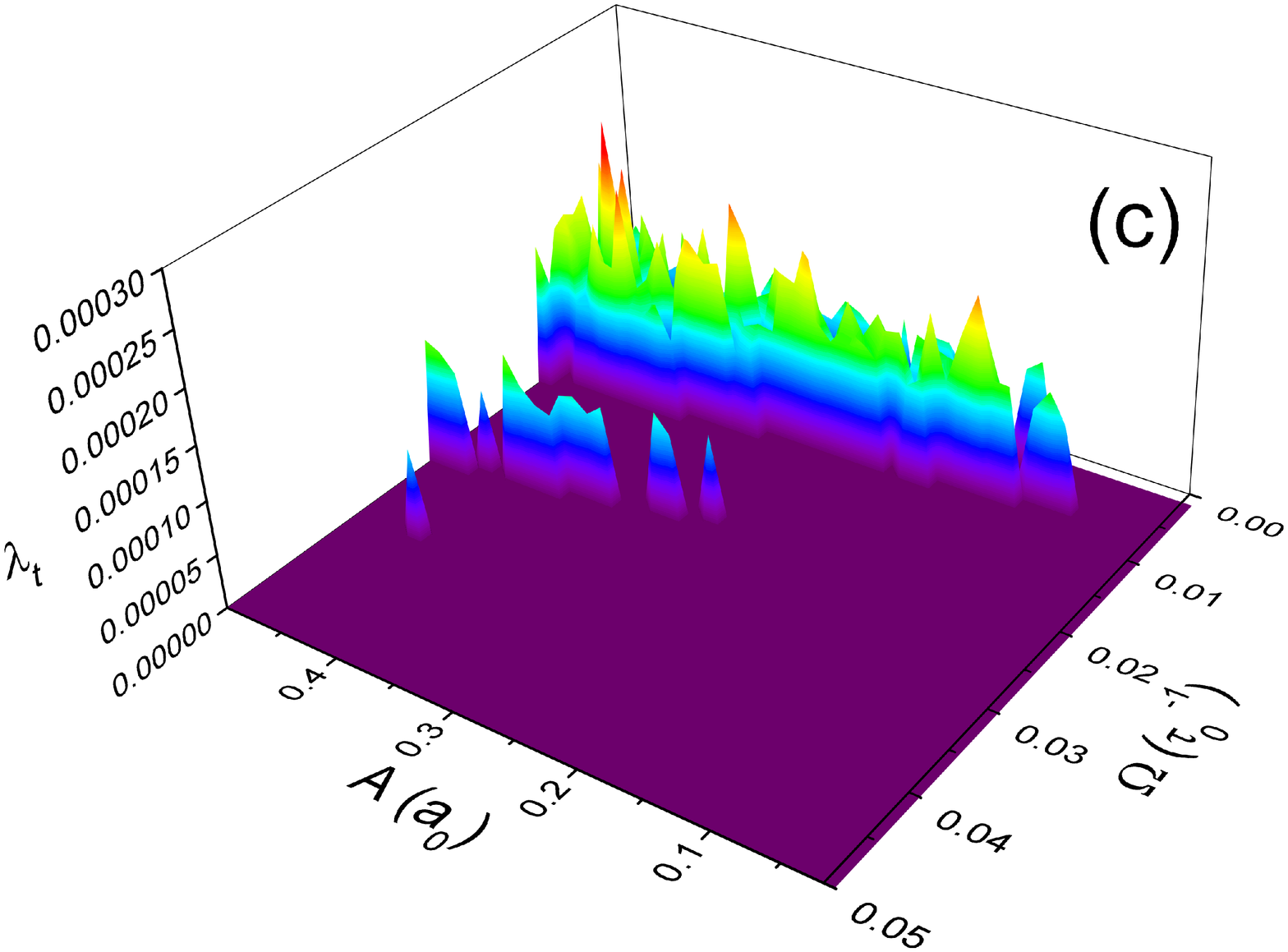}
\caption{Values of the Lyapunov exponents $\lambda_{R}$, $\lambda_{\varepsilon}$, and $\lambda_{t}$ in the $A$-$\Omega$ parameter space. 
         The case of ${\rm H^{+}_{2}}$ cation (the Gaussian orbitals).}
\label{f03III}
\end{figure*}

Variationally computed dependence of the total energy $E_{T}$ on the distance $R$ models the effective potential of interaction between atomic cores in the cation. If the cores are additionally influenced by the harmonic force of the amplitude $A$ and the frequency $\Omega$, then the Newton equation which determines the $R\left(T\right)$ function takes the form: 
\begin{eqnarray}
\label{r01-III}
\mu\frac{d^{2}R\left(T\right)}{dT^{2}}=-\left[\frac{dE_{T}\left(r\right)}{dr}\right]_{r=R\left(T\right)}-A|\cos\left(\Omega T\right)|,
\end{eqnarray}
where the quantity $\mu=M_{C1}M_{C2}/\left(M_{C1}+M_{C2}\right)$ denotes the reduced mass of atomic cores. The minimum value of $\mu$ is obtained for the ${\rm H_{2}^{+}}$ cation and is equal to $918.076336$ (in electron mass units $m_{e}$). 
For more complicated systems, the mass of the given core $M_{C}$ can be estimated from the formula:
$M_{C}\sim \left(n_{p}+n_{n}\right)m_{p}$, where $n_{p}$ ($n_{n}$) represents the number of protons (and neutrons) contained in this core, $m_{p}$ is the proton rest mass. We neglected the contribution from the intercore electrons because $m_{e}\ll m_{p}$.

It should be mentioned that we take into account only the case when the external harmonic force stretches the molecule. Therefore the absolute value symbol occurs in equation \eq{r01-III}. We do not analyse the case of the compressive force to exclude the possibility of the molecule rotation, which could proceed perpendicularly to the direction of the external force.  This convenient limitation refers to the fact that the fundamental properties of the atomic core dynamics (e.g. the chaotic state occurrence) do not depend on the exciting force direction (its inward or outward orientation), but result from the strong nonlinearity of the effective intercore potential \cite{Schuster1984A, Jarosik2018A}.  

As we already know the total energy function $E_{T}\left(R\right)$ calculated from the first principles, we can solve the Newton equation. On the account of our interest in the chaotic behaviour of the molecule, we presented the adequately selected results for the intercore distance versus time in Figure \fig{f01III}~(a). Precisely speaking, we took into account two trajectories, $R_{1}\left(T\right)$ and $R_{2}\left(T\right)$, which initially were $10^{-6}$~${\rm a_{0}}$ apart. Additionally, we assumed that the value of Lyapunov time ($\mathcal{T}$) would be reached when the trajectories went apart as far as the distance of $10^{-1}$~${\rm a_{0}}$. The Lyapunov time for the case presented in Figure \fig{f01III}~(a) occurred to be $\mathcal{T}_{R}=9664.4$~${\rm\tau_{0}}$, where $\tau_{0}\simeq 2.418\cdot 10^{- 17}$~s is the unit of time in the atomic system of units.

The time dependence of the intercore distance correlates directly to the time dependence of electronic Hamiltonian parameters (the energy of the electron orbital $\varepsilon$ and the hopping integral $t$). This results from the fact that these quantities depend directly from the distance $R$ (see \mbox{Figure \fig{f02II}}).  
Figures \fig{f01III}~(b) and (c) represent the time dependence of the energy of the electron orbital $\varepsilon$ and the hopping integral $t$. The obtained trajectories also indicate the chaotic behaviour, characterised with Lyapunov time values $\mathcal{T}_{\varepsilon}=10973.2$~${\rm\tau_{0}}$ and $\mathcal{T}_{t}=10973.2$~${\rm\tau_{0}}$, respectively. 
The values of $\mathcal{T}_{\varepsilon}$ and $\mathcal{T}_{t}$ are quite close to the value of $\mathcal{T}_{R}$, what signifies that these parameters can be used interchangeably to characterise the chaotic behaviour of ${\rm H^{+}_{2}}$ system.  

Besides the Lyapunov time, also the structure of the power spectrum 
$P_{x}\left(\omega\right)=|\lim_{a\rightarrow +\infty}\int^{a}_{0}dT\exp\left(i\omega T\right)x\left(T\right)|^2$, where $x=\varepsilon$ or $x=t$, gives the evidence of chaos. Figures \fig{f02III}~(a) and (b) show our results in this respect. One can see the large broadening of the power spectrum indicating the great number (theoretically an infinite number) of frequencies present in examined signals. It is the characteristic feature of the power spectra of functions changing in the chaotic way (irregular and non-periodical). Let us notice that for the quasi-periodic functions, which are also very complicated, one obtains the discrete lines corresponding to the definite frequencies \cite{Schuster1984A}.

The overall characteristics of the chaotic properties of ${\rm H_{2}^{+}}$ cation is presented in diagrams \fig{f03III}~(a)-(c), where we plotted the values of the Lyapunov exponents \mbox{($\lambda\sim 1/\mathcal{T}$)}, namely $\lambda_{R}$, $\lambda_{\varepsilon}$, and $\lambda_{t}$, versus the amplitude $A$ and the frequency $\Omega$ of the exciting force. It can be easily seen in the Figure \fig{f03III}~(a) that the atomic core subsystem is in the chaotic state only for selected values of $A$ and $\Omega$. Values of the $\lambda_{R}$ exponent corresponding to the chaotic state form characteristic 'islands', the one particularly extensive being found for low exciting frequencies. The juxtaposed diagrams \fig{f03III}~(b) and (c), presenting $\lambda_{\varepsilon}$ and $\lambda_{t}$, are very similar to the diagram $\lambda_{R}\left(A,\Omega\right)$, as was expected.

The presented model allows also to analyse the chaotic behaviour of the ${\rm XY^{+}}$-type cations, where the molecular bonding is realised by means of the single electron. Such cations can differ from the ${\rm H_{2}^{+}}$ cation with respect to the reduced mass $\mu$ and/or to the value of the $|\Delta Z|$ parameter. Our results prove that the chaotic behaviour of the considered group of cations has similar characteristics as 
the ${\rm H_{2}^{+}}$ ion.  

%%%%%%%%%%%%%%%%%%%%%%%%%%%%%%%%%%%%%%%%%%%%%%%%%%%%%%%%%%%%%%%%%%%%%%%%%%%%%%%%%%%%%%%%%%%%%%%%%%%%%%%%%%%%%%%%%%%%%%%%%%%%%%%%%%%%%%%%%%%%%%%%%%%%%%%%(V)
\section{SUMMARY AND DISCUSSION OF RESULTS}

In the paper, we studied the dynamic properties of the molecules 
${\rm X_{2}^{+}}$ (${\rm H_{2}^{+}}$) and ${\rm XY^{+}}$ ($\rm LiH^{+}$, $\rm LiNa^{+}$) both with symmetric and the asymmetric charge distribution. In the case when the cations are subjected to vibration by the harmonic force, the chaotic changes of the intercore distance $R$ can be observed for some values of the force amplitude $A$ and the force frequency $\Omega$. It should be emphasized that the chaotic behaviour of the examined systems results from the presence of the highly non-linear intercore potential, and not from the specific form of the exciting force. The comprehensive analysis of the atomic core dynamics shown that there exist characteristic areas in the $A$-$\Omega$ parameter space for which 
the non-zero values of Lyapunov exponent $\lambda_{R}$ can be found. The particularly large 'islands' of this type were revealed for the low excitation frequencies.

The chaotic changes of the intercore distance induce directly the chaotic evolution of the electronic Hamiltonian parameters 
$\varepsilon$ and $t$. It should be stressed, that the change of the core mass or the charge distribution not influences qualitatively the general structure of the diagrams $\lambda_{R}\left(A,\Omega\right)$ or $\lambda_{x}\left(A,\Omega\right)$, where 
$x=\varepsilon$ or $x=t$. Moreover, the great similarity between $\lambda_{R}\left(A,\Omega\right)$ and $\lambda_{x}\left(A,\Omega\right)$ can be observed.

The results presented in the paper were obtained as part of the formalism of the second quantization using the variational calculations. Our method of analysis is one of the most accurate methods that can be used. This is evidenced by the results that we received for ${\rm H_ {2}^{+}}$, which turned out to be consistent with the results obtained by Schaad and Hicks \cite{Schaad1970A}. It should be emphasized that the discussed method can using in order to successfully characterized even more complex systems than single-electron cations. In particular, for the hydrogen molecule 
${\rm H_{2}}$ and the anion ${\rm H^{-}_{2}}$, it has been obtained:
$R_{0}=1.41968$~${\rm a_ {0}}$, ($E_ {0}=-2.323011$~Ry), and 
$R_{0}=3.476828$~${\rm a_{0 }}$, ($E_{0}=-1.947958$~Ry). 
In the case of the hydrogen molecule, our results are fully consistent with the Ko{\l}os and Wolniewicz data \cite{Kolos1964A, Kolos1968A}:
$R_{0}=1.3984$~${\rm a_{0}}$ and $E_{0}=-2.349$~Ry, and the results obtained by K{\c{a}}dzielawa 
{\it et al.} \cite{Kadzielawa2014A}: $R_{0}=1.43042$~${\rm a_{0}}$ and $E_{0}=-2.29587$~Ry. 
Please note that, the value of $E_{0}$ calculated using the {\it Mopac} software package \cite{Mopac} 
differs from the results of Ko{\l}os and Wolniewicz by $12$~\%. The calculations made by us using 
the {\it Quantum Espresso} \cite{Baroni1986A} package gave the inaccurate value of the dissociation energy ($\sim 0.17$~Ry, the PBE functional). 
For the ion ${\rm H_{2}^{-}}$, literature data are divergent. The early theoretical paper by Eyring, Hirschfelder, and Taylor, using the valence bond technique with two variation parameters, found the stable ground state with the minimum at $R_{0}=3.40151$~${\rm a_ {0}}$ \cite{Eyring1936A}. This result correlates well with ours, where $R_{0}=3.476828$~${\rm a_{0}}$. However, the paper \cite{Taylor1963A} suggests that the ion ${\rm H_{2}^{-}}$ is not stable relative to the auto-ionization into ${\rm H_{2}}$ and the electron at infinity.

Nevertheless, the presented model contains some simplifications. 
In particular, we use the Born-Oppenheimer approximation 
\cite{Born1927A}, and we treat the atomic nucleus with the inner shell electrons as the atomic core. 
Using the Born-Oppenheimer approximation in our calculations did not include the non-adiabatic terms that would give better approximations when expanded to orders not less than $r^{1/4}$, where $r\sim 544\cdot 10^{-6}$ is the ratio of the electron to the proton mass \cite{Born1954A}. On the other hand, the atomic core approximation is commonly used in quantum chemistry, allowing to obtain the correct results for the right screening constant $Z$. 
It should be clearly emphasized that the calculations carried out outside the Born-Oppenheimer and atomic core approximation will not significantly change our results regarding the chaotic evolution of electron parameters $\varepsilon$ and $t$, because they will not lead to the linear relationship between total energy ($E_{T}$) and the distance $R$ between the atoms. In our view will be the opposite, the $E_{T}\left(R\right)$ function should contain additional small nonlinear corrections.

The results obtained in the paper should be of particular interest to researchers dealing with the analysis of dynamics of open systems far from equilibrium, where the strong correlations and the nonlinear effects occur simultaneously \cite{Balzer2013A}. The out-of-equilibrium dynamics 
are of great current interest in the molecular physics (molecules \cite{Wrona2019A} or the Hubbard nanoclusters \cite{Hermanns2014A}), 
in the solid-state physics \cite{Eckstein2009A, Eckstein2010A, Moeckel2010A}, in the optical lattices \cite{Bloch2008A}, and in the quantum transport \cite{Uimonen2011A, Khosravi2012A, Khosravi2014A}. In all these cases, due to the highly non-linear interatomic potentials, for properly selected excitation, the dynamics of electron parameters should be chaotic. Please note that currently such research can be successfully carried out because, we have very advanced tools for analyzing the complex dynamics of quantum systems. Let's list here the exact diagonalization \cite{Acquarone1998A, Kadzielawa2014A}, the density matrix renormalization group approaches \cite{White1992A, Kennes2016A, Mitra2017A}, the nonequilibrium dynamical mean field theory \cite{Eckstein2009A, Tsuji2014A} and iterative path integral \cite{Weiss2013A}. What is particularly important theoretically obtained results can be confronted with experimental results (the time-resolved spectroscopy experiments 
\cite{Ogasawara2000A, Perfetti2006A, Wall2011A}, and experiments on ultracold atoms trapped in the optical lattice \cite{Bloch2008A, Schneider2008A}).

The molecular stability issues are another compelling research area for which our results may be useful. 
It can be easily seen that the molecule subjected to external excitation showing chaotic changes in atomic distances will be very susceptible to dissociation \cite{Jarosik2018A}. This effect can be eliminated or strengthened by appropriately selecting the parameters of excitation. Possible applications range from chemistry to molecular biology.

%%%%%%%%%%%%%%%%%%%%%%%%%%%%%%%%%%%%%%%%%%%%%%%%%%%%%%%%%%%%%%%%%%%%%%%%%%%%%%%%%%%%%%%%%%%%%%%%%%%%%%%%%%%%%%%%%%%%%%%%%%%%%%%%%%%%%%%%%%%%%%%%%%%%%%%(VI)
%
\begin{acknowledgements}
J. K. K. wishes to thank the ERDF/ESF project 'Nanotechnologies for Future' (CZ.02.1.01/0.0/0.0/16\_019/0000754) for the financial support.
J. K. K. also acknowledges the financial support from the program of the Polish Minister of Science and Higher Education under the name ``Regional Initiative of Excellence'' in 2019-2022, project no. 003/RID/2018/19, funding amount 11 936 596.10 PLN.
\end{acknowledgements}
% 

%%%%%%%%%%%%%%%%%%%%%%%%%%%%%%%%%%%%%%%%%%%%%%%%%%%%%%%%%%%%%%%%%%%%%%%%%%%%%%%%%%%%%%%%%%%%%%%%%%%%%%%%%%%%%%%%%%%%%%%%%%%%%%%%%%%%%%%%%%%%%%%%%%%%%%(VII)
\bibliography{Bibliography}
%%%%%%%%%%%%%%%%%%%%%%%%%%%%%%%%%%%%%%%%%%%%%%%%%%%%%%%%%%%%%%%%%%%%%%%%%%%%%%%%%%%%%%%%%%%%%%%%%%%%%%%%%%%%%%%%%%%%%%%%%%%%%%%%%%%%%%%%%%%%%%%%%%%%%%%%%%%
\end{document}